\documentstyle[12pt]{article}

\makeatletter                                                           

\def\section{\@startsection {section}{1}{\z@}{-1.5ex plus -.5ex         
minus -.2ex}{1ex plus .2ex}{\large\bf}}                                 

\def\@thmcountersep{}                                                   

\long\def\@makecaption#1#2{\vskip 10pt \setbox\@tempboxa\hbox{#1. #2}   
   \ifdim \wd\@tempboxa >\hsize   
       #1. #2\par                 
     \else                        
       \hbox to\hsize{\hfil\box\@tempboxa\hfil}                         
   \fi}                                                                 

\makeatother                                                            

\title{An Illustration of 2+1 Gravity Loop Transform Troubles}

\author{
Donald Marolf \thanks{Physics Department, Syracuse University, Syracuse,
NY 13244-1130, USA. This author was supported in part by NSF grant
PHY-9005790 and by research funds provided by Syracuse University.}}

\date{} 

\begin{document}
\maketitle
\pagestyle{headings}                                                   
\flushbottom                                                           

\vspace{-10pt} 

{\centerline{SU-GP-93/3-4, gr-qc/9305015}}
{\centerline{For the proceedings of the 5th Canadian Conference on
General Relativity and Astrophysics}}
\begin{abstract}

A nonperturbative approach to
quantum gravity that has generated much discussion is
the attempt to construct a ``loop representation."
Despite it's success in linear
quantum theories and a part of
2+1 quantum gravity, it has recently
been noticed that difficulties arise with loop
representations in a different ``sector" of 2+1
gravity.  The problems are related to the use of the ``loop transform"
in the construction of the loop representation.  We
illustrate these difficulties
by exploring an analogy based on the Mellin transform which allows
us to work in a context that is
both mathematically and physically simple and that does not
require an understanding either of loop representations or of 2+1
gravity.
\end{abstract}

\def\ie{{i.e.}}
\def\eg{{e.g.}}
\def\three{\,^3 \kern-1.5pt}

\def\newblock{}
\newcommand{\half}{\frac{1}{2}}

\def\pmb#1{\setbox0=\hbox{$#1$}%
  \kern-.025em\copy0\kern-\wd0
  \kern.05em\copy0\kern-\wd0
  \kern-.025em\raise.0433em\box0}
\def\pmbs#1{\setbox0=\hbox{$\scriptstyle #1$}%
  \kern-.0175em\copy0\kern-\wd0
  \kern.035em\copy0\kern-\wd0
  \kern-.0175em\raise.0303em\box0}

\def\bfbeta{{\pmb{\beta}}}
\def\bfsbeta{\pmbs{\beta}}

\def\fraction#1#2{{\textstyle{#1\over#2}}} \def\fr{\fraction}
\def\sfraction#1#2{{\scriptstyle{#1\over#2}}} \def\sfr{\sfraction}
\def\d{\partial}

\def\vol#1{{\bf #1}}
\def\journalfont{\it}         
\def\jou#1{{\journalfont #1\ }}
\def\aaa{\jou{  Astron.\ Astrophys.}}
\def\aip{\jou{  Adv.\ Phys.}}
\def\am{\jou{   Ann.\ Math.)}}
\def\ap{\jou{   Ann.\ Phys.\ (N.Y.)}}
\def\apj{\jou{  Astrophys.\ J.}}
\def\cjp{\jou{  Can. J. Phys.}}
\def\cmp{\jou{  Commun.\ Math.\ Phys.}}
\def\cqg{\jou{  Class.\ Quantum Grav.}}
\def\grg{\jou{  Gen.\ Relativ.\ Grav.}}
\def\jmp{\jou{  J.\ Math.\ Phys.}}
\def\jpamg{\jou{ J.\ Phys.\ A: Math.\ Gen.}}
\def\mnras{\jou{ Mon.\ Not.\ R.\ Ast.\ Soc.}}
\def\nat{\jou{  Nature}}
\def\ncim{\jou{ Nuovo Cim.}}
\def\nucp{\jou{ Nuc.\ Phys.}}
\def\ncb{\jou{  Il Nuovo Cimento ``B}}
\def\pl{\jou{   Phys.\ Lett.}}
\def\pr{\jou{   Phys.\ Rev.}}
\def\prep{\jou{ Phys.\ Rep.}}
\def\prl{\jou{  Phys.\ Rev.\ Lett.}}
\def\ptp{\jou{  Prog.\ Theor.\ Phys.}}
\def\rmp{\jou{  Rev. Mod. Phys.}}
\def\spj{\jou{  Sov.\ Phys.\ JETP}}
\def\gr{04.20.-q}            
\def\cosmology{98.80.Dr}     

\def\Lie{{\cal L}}
\def\Mscr{{\cal M}}
\def\halfcone{\hbox{$\Mscr^-_{1/2}$}}
\def\cone{\hbox{$\Mscr^-$}}
\def\IP#1#2{\langle\, #1\, |\, #2\, \rangle}  
\def\Div#1#2{{\rm Div}_{#1}#2}                
\def\cV{{\cal V}}
\def\ps{phase space}
\def\qtzn{quantization}
\def\dd{{\rm d}}
\def\Real{{\rm I}\! {\rm R}}

\newcommand{\bu}{\bar u}
\newcommand{\bz}{\beta^0}
\newcommand{\bp}{\beta^+}
\newcommand{\bm}{\beta^-}
\newcommand{\bbz}{\bar\beta^0}
\newcommand{\bbp}{\bar\beta^+}
\newcommand{\bbm}{\bar\beta^-}
\newcommand{\tphi}{\tilde\phi}
\newcommand{\tbz}{\tilde\beta^0}
\newcommand{\tbp}{\tilde\beta^+}
\newcommand{\tbm}{\tilde\beta^-}
\newcommand{\tp}{\tilde{p}}
\newcommand{\tpz}{{\tilde p}_0}
\newcommand{\tpp}{{\tilde p}_+}
\newcommand{\tpm}{{\tilde p}_-}
\newcommand{\bpz}{{\bar p}_0}
\newcommand{\bpp}{\bar{p}_+}
\newcommand{\bpm}{\bar{p}_-}

\newcommand{\be}{\begin{equation}}
\newcommand{\ee}{\end{equation}}



\catcode`@=11

\newdimen\jot \jot=3pt
\newskip\z@skip \z@skip=0pt plus0pt minus0pt
\newdimen\z@ \z@=0pt 
\dimendef\dimen@=0

\def\m@th{\mathsurround=\z@}

\def\ialign{\everycr{}\tabskip\z@skip\halign} 

\def\openup{\afterassignment\@penup\dimen@=}
\def\@penup{\advance\lineskip\dimen@
  \advance\baselineskip\dimen@
  \advance\lineskiplimit\dimen@}

\def\eqalign#1{{
\null\,\vcenter{\openup\jot\m@th
  \ialign{\strut\hfil$\displaystyle{##}$&$\displaystyle{{}##}$\hfil
      \crcr#1\crcr}}\,} }
\def\meqalign#1{\null\,\vcenter{\openup\jot\m@th
  \ialign{\strut\hfil$\displaystyle{##}$&&$\displaystyle{{}##}$\hfil
      \crcr#1\crcr}}\,}
\catcode`@=12   

\section{Introduction}

Because
no satisfactory perturbative theory has been found, nonperturbative
approaches to
quantum gravity are growing in popularity.  One such approach \cite{SD1}
uses a canonical framework and is base on
self-dual connections instead of metrics.  Our discussion will be concerned
with a particular viewpoint within this appraoch.

Because the fundamental objects of this theory are connections, holonomies
around closed loops and their traces provide an important step toward a gauge
invariant description.  Loops are then a part of the corresponding
formulations of
quantum gravity as well.
One idea for quantum gravity takes these loops
very seriously and attempts to formulate the theory in terms of
functions of loops by using a set of such functions to carry a
representation of a ``loop algebra," thus creating a
a ``loop representation" \cite{Ash}.  Much effort has gone into
this study \cite{Ash,Rov}
but a loop description of quantum gravity is far
from completed.  For this reason, simple systems for which we can
thoroughly analyze loop representations are important and such
systems have been studied with great success.  Examples include
linear systems and
2+1 dimensional gravity on ${\cal T}^2 \times {\cal R}$ (see
references in \cite{long}). Until recently, however, only a
part of the 2+1 gravity model with a compact gauge group
had been studied in these terms.  The groups in the linear models
are also compact, in contrast to 3+1 gravity.

Ref. \cite{long} explores loop representations of another ``sector"
of 2+1 gravity and finds quite different results.
Because the gauge group in this new sector is not compact,
the ``loop transform" integral used to define the loop representation
is not well behaved and cannot be used naively.
A loop representation can still be constructed,
though it is not entirely satisfactory.

The loop transform in 2+1 quantum gravity is a link between
two representations of the so-called ``loop algebra."  This
is an algebra of quantum operators labelled by ``loops,"
where ``loops" in the 2+1 torus case means elements of $\pi_1({\cal T}^2)$ --
essentially pairs of integers.  One
representation
is carried by ${\cal L}^2({\cal R}^2)$ on which
the operators act by multiplication and differentiation.  The other is
the ``loop representation" in which the operators act by
multiplication and translation on functions of pairs of integers.
The loop transform maps an element of ${\cal L}^2({\cal R}^
2)$ to a function of pairs of integers and preserves the action of
the loop operators.

Because the ${\cal L}^2$
representation is related to reduced phase space quantization, it always has
certain properties (such as conditions on the spectra of the fundamental
operators) that help to produce the correct classical limit.  On the
other hand, the loop representation does not in general have these properties.
When the loop transform is an isomorphism it guarantees that the
proper conditions hold in the
loop representation as well, but when it is not special
steps must be taken.

Technicalities make a discussion of the 2+1 loop transform and
loop representations complicated, but there is a strong analogy with a simple
model that uses the Mellin transform instead.  It is this model that we
will turn to now so that we can illustrate the problems involved with
minimal technical complications.

\section{The Mellin Model}
\label{MM}

In this new sector of 2+1 quantum gravity, the loop
transform $L$ is just the Laplace transform:
\begin{equation}
(L \psi)(n) = \int d^2x \ e^{n_1x_1 + n_2x_2} \psi(x_1,x_2)
\end{equation}
evaluated at integer values, $n$.  Two characteristics of this transform
stand out:  first, the transform does not converge for
all $\psi \in {\cal L}^2
({\cal R})$; and second, the transform is evaluated only at integer values.
Another transform with these characteristics is the integer Mellin
transform:
\begin{equation}
\label{Mellin}
\hat{\psi}(n) \equiv \int_{-\infty}^{\infty} x^n \psi (x) dx
\end{equation}
for nonnegative integers $n$.  As with the loop transform, we can also think
of the Mellin transform
as a link between two representations of an operator algebra: one carried by
${\cal L}^2({\cal R})$ and one carried by functions of nonnegative
integers.  The algebra in question contains operators $\{X_n,D_m\}$
for nonnegative integers $n$ and $m$ and is defined by the commutation
relations:

\begin{equation}
\label{Melcomms}
[X_n, X_m] = 0, \qquad [X_n, D_m] = inX_{(n+m)}, \qquad
[D_n, D_m] = -i(m-n)D_{m+n}
\end{equation}
The operators $X_n = x^n$ and
$D_m = -ix^{(m+1)} {{\partial} \over {\partial x}}
-i {{n+1} \over {2}} x^m$ form a representation of this algebra on ${\cal
L}^2({\cal R})$ that is Hermitian with respect to the
the  ${\cal L}^2({\cal R})$
inner product.  The representation on
functions of nonnegative integers $\hat{\psi}(n)$ is given by:

\begin{equation}
X_k \hat{\psi}(n) = \hat{\psi}(n+k) \ {\rm and} \ D_m\hat{\psi}(n) =
i (m + {{n+1} \over {2}}) \hat{\psi}(n+m)
\end{equation}
The Mellin transform maps functions on ${\cal R}$ to functions on
${\cal Z}^+ \cup \{0\}$ and maps the action of our operators
on ${\cal L}^2({\cal R})$ to their action on functions of integers. The
difficulties of using the Mellin transform become clear in the momentum
representation:

\begin{equation}
\hat{\psi}(n) = \int x^n
\tilde{\psi}(p) e^{-ixp}dxdp = 2 \pi \bigl( -i
{{\partial} \over {\partial p}}  \bigr) ^n \tilde{\psi}(p) \bigg|_{p=0}
\end{equation}
The function $\hat{\psi}(n)$ determines
$\psi(x)$ uniquely {\em only} if we require that $\tilde{\psi}(p)$ be
analytic and the transform annihilates any function
whose Fourier transform vanishes in some neighborhood of
the origin.  Thus, our transform annihilates a dense subspace of
the ${\cal L}^2$ space and is not
an isomorphism between the ${\cal L}^2$ representation and {\em any}
representation carried by functions of integers.
This is just what \cite{long} shows
for the loop
transform.

However, there are also several
subspaces of ${\cal L}^2$ that: i) are dense,
ii) provide a representation of the
above algebra,
iii) lie inside the transform's domain, but
iv) have trivial intersection with the transform's kernel.
One example is $S = \{P(x,e^{x^2}) \exp(1-e^{x^2}):
P(x)$ is a polynomial in $x$ and $e^{x^2}\}$ which we
now use to build an ``integer function representation," despite
the transform's dense kernel.

The construction proceeds as follows:  First, we map
each element of $S$ to a function of integers as in Eq. \ref{Mellin}.
The exact result is difficult to
write in a useful form, but we will be content with a crude approximation:
note that $\exp(kx^2+1-e^{x^2})$ falls off sharply
when $e^{x^2}-1-kx^2 \approx 1$ and replace it with a step function.
If we write $f_{m,k}(x) = x^m \exp(kx^2+1-e^{x^2})$, then

\begin{eqnarray}
g_{m,k}(n) \equiv  (Mf_{m,k})(n) &=& \int_{-\infty}^{\infty} dx \ x^{n+m}\
\exp(kx^2+1-e^{x^2})
\cr &\approx& \int_{-\lambda_k}^{\lambda_k} dx
\ x^{n+m} = 2{\lambda_k^{n+m+1} \over {n+m+1}}
\end{eqnarray}
where $\lambda_k$ is the positive root of $e^{\lambda_k^2}-1-k\lambda_k^2
= 1$ and
$M$ denotes the Mellin transform.
Although this is only an approximation, it is
clear that the functions $\{Mf_{k,m}\}$ are linearly independent so that
$M$ annihilates only the zero element of $S$.

We now define the inner product on this
image space using the inner product on the ${\cal L}^2$
space.  For $g_1,g_2 \in M(S)$, let $\langle g_1,g_2 \rangle
= \langle M^{-1}g_1,M^{-1}g_2 \rangle$.  Note that $M^{-1}$ is
well-defined as a map from $M(S)$ to $S$ since $S$ has trivial intersection
with the kernel of $M$.  Roughly,
The result is summarized by:
\begin{equation}
\label{norm}
\langle g_{k,m},g_{l,n} \rangle \approx 2 {(\gamma_{k+l})^{n+m+1} \over
{n+m+1}}
\end{equation}
using the same crude approximation as before and where $\gamma_q$
satisfies $2e^{\gamma_q^2}-2-q\gamma^2_q = 1$.

We then complete the image space $M(S)$ with respect to this inner
product.  Since $S$ was dense in the ${\cal L}^2$ space and we have used the
same inner product on $S$ and $M(S)$, this completion will be isomorphic to
the ${\cal L}^2$ space.  It is in this sense that we can construct an
integer function representation that is
isomorphic to the ${\cal L}^2$ representation.

The trouble is that not all elements of this completed space are functions of
integers, at least not in a useful way.  Consider for a moment the more
familiar $l^2$ inner product on functions of integers.  If a sequence of such
functions converges in $l^2$, then it always converges to a function of
integers.  In fact, it converges to that function pointwise.  The point is
that the inner product defined by Eq. \ref{norm} is not so pleasant.  Note
that since this inner product is defined only on functions in the image
$M(S)$ it is difficult to say in general whether ``a sequence $\{Ms_n\} \subset
M(S)$ that converges actually converges to a function of integers"
unless that limit function is in $M(S)$ as well.  Suppose
then that we consider $\{s_n\} \subset {\cal L}^2$ that converges to some
$\tilde{s} = s+k$ for some $s \in S$ and some $k$ in the kernel of $M$.  Then
$M\tilde{s} \in M(S)$ so that it is meaningful to ask if $\{Ms_n\}$ converges
to $M\tilde{s}$.  Note, however, that by construction
\begin{eqnarray}
\lim_{n \rightarrow \infty} ||Ms_n - M\tilde{s}||^2 &=& \lim_{n \rightarrow
\infty} ||Ms_n - Ms||^2 = \lim_{n \rightarrow \infty} ||s_n - s||^2 \cr
&=& ||\tilde{s} - s||^2 = ||k||^2 \neq 0
\end{eqnarray}
so that $MS_n$ does not converge to $M\tilde{s}$ and certainly doesn't
converge to any other function in $M(S)$.  We might choose to somehow define
the sequence to converge to some function outside of $M(S)$ and to define an
extension of the inner product that makes this consistent, but this would have
little meaning.

Another problem is that $S$ is not the only subspace that satisfies (i)-(iv).
Note that the space we have been using (or one that would
work just as well) can be obtained by
starting with the single function $\exp(1-e^{x^2})$ and applying all possible
combinations of $X_n$'s and $D_m$'s.  However, we could also start
with any function of the form $\exp(1-e^{x^2}) + K(x)$ for some $K(x)$
in the kernel of $M$ and again obtain a subspace $S_2$ satisfying
(i)-(iv) by applying $X_n$'s and $D_m$'s.  Properties (i)-(iii)
are straightforward and we can show (iv) by studying the
function $\exp(1-e^{x^2})$.

To do so, first note that the operators $X_n$ and $D_m$ preserve the
kernel of $M$.  This means that every function in $S_2$
is a sum of a function in our original set $S$
and a member of the kernel.  Since $S$ has trivial
intersection with this kernel, the only way that $S_2$ can
have nontrivial intersection is if some combination
of $X_n$'s and $D_m$'s annihilates
$\exp(1-e^{x^2})$ but not $K(x)$.

Now, recall that $X_n$ and $D_m$ act on ${\cal L}^2({\cal R})$
by taking derivatives
and multiplying by polynomials in $x$.  A short calculation shows
that the derivatives of $\exp(1-e^x)$ are of the form:
\begin{equation}
({{\partial} \over {\partial x}})^n  \exp(1-e^{x^2}) =  \exp(1-e^{x^2})
\sum_{k=0, l=0}^n \alpha_{k,l}^n x^l e^{kx^2}
\end{equation}
with positive real coefficients $\alpha_{k,l}^n$ such that $\alpha^n_{n,0} \neq
0$.  Even
if we multiply these expressions by arbitrary polynomials, no sum of
such terms can be zero and no combination of $X_n$ and $D_m$ can
annihilate $\exp(1-e^{x^2})$!  Our new set $S_2$ works just fine.  Note
that the image $M(S_2)$ is exactly the same as the image of
our first set $S$.

It is therefore unclear just which elements of ${\cal L}^2$ the
functions $g_{m,k}(n)$
represent -- each could come from an element of either $S$ or $S_2$.
Even the inner product of these functions depends on which set we
choose, as we can see by making the following choice of $K(x)$.
Since the
kernel is dense in ${\cal L}^2({\cal R})$, we can choose $K(x)$ arbitrarily
close to $-f_{0,0}$ so that the set $S_2$ leads to the
inner product: \begin{eqnarray}
\langle g_{k,m},g_{l,n} \rangle &\approx& \langle f_{k,m}-f_{0,0},
f_{l,n} - f_{0,0} \rangle \cr &\approx& \langle f_{k,m},f_{l,n} \rangle
- \langle f_{k,m},f_{0,0} \rangle - \langle f_{0,0},f_{l,n} \rangle
+ \langle f_{0,0},f_{0,0} \rangle \cr &\approx&
2[{( \gamma_{k+l})^{n+m+1} \over {n+m+1}} - {(\gamma_{k})^{m+1} \over {m+1}}
- {(\gamma_{l})^{m+1} \over {n+1}} + ln(3/2)]
\end{eqnarray}
Note that the norm of  $g_{0,0}(n)$ is almost zero
while the norms of other functions are still quite large.  The set $S_2$
therefore leads to a genuinely different inner product
on the functions of integers and not just a rescaling of the one
defined by $S$.

This is the kind of difficulty that researchers of loop representations
will have to face.  There may well be quantum theories
with reasonable properties that {\em could} be defined in terms of loops, but
if 2+1 gravity and the Mellin transform are any guide,
their construction must use some extra structure to
link the loop description to a more familiar one.  Furthermore,
this description will be
highly dependent on this choice and the structure will be difficult to
discard after it has been used.  This suggests
that there is no way to avoid discussing difficult and
fundamental questions such as the choice of a Hilbert space
in terms of some more familiar approach to quantum gravity.



\begin{thebibliography}{1}


\bibitem{SD1}  Ashtekar A 1987 {\it Phys. Rev. D} {\bf 36} 1587-603
\bibitem{Ash}  Ashtekar A 1991 {\it Lectures on Nonperturbative Canonical
Gravity} (Singapore: World Scientific)
\bibitem{Rov}  Rovelli C 1991 {\it Class. Quant. Grav.} {\bf 8} 1613
\bibitem{long} Marolf D 1993 {\it preprint} Syracuse University
 SU-GP-93/3-1

%
\end{thebibliography}

\bibliographystyle{plain}

\end{document}